\documentclass[%
 reprint,
superscriptaddress,
 amsmath,amssymb,
 aps,
]{revtex4-1}

\usepackage{graphicx}% Include figure files
\usepackage{dcolumn}% Align table columns on decimal point
\usepackage{bm}% bold math
\usepackage{braket}
\usepackage[utf8]{inputenc} % for ö,ä,ü...
\usepackage{upgreek}
\usepackage{gensymb}
\usepackage{float}

\newcommand{\unit}[1]{\,\text{#1}}

\newcommand{\Evs}{E_{\text{VS}}}

\usepackage[colorlinks=true,linkcolor=blue]{hyperref}%% add hypertext capabilities
\usepackage{subcaption}
\begin{document}

%\preprint{APS/123-QED}

\title{Large, tunable valley splitting and single-spin relaxation mechanisms\\ in a Si/Si$_x$Ge$_{1-x}$ quantum dot}% Force line breaks with \\

\author{Arne Hollmann}%
%\email[email to: ]{arne.hollmann@rwth-aachen.de}
\affiliation{JARA-FIT Institute for Quantum Information, Forschungszentrum J\"ulich GmbH and RWTH Aachen University, Aachen, Germany}
%\date{August 10, 2010}%
\author{Tom Struck}%
\affiliation{JARA-FIT Institute for Quantum Information, Forschungszentrum J\"ulich GmbH and RWTH Aachen University, Aachen, Germany}
\author{Veit Langrock}%
\affiliation{JARA-FIT Institute for Quantum Information, Forschungszentrum J\"ulich GmbH and RWTH Aachen University, Aachen, Germany}
\author{Andreas Schmidbauer}%
\affiliation{Institut für Experimentelle und Angewandte Physik, Universit\"at Regensburg, Regensburg, Germany}
\author{Floyd Schauer}%
\affiliation{Institut für Experimentelle und Angewandte Physik, Universit\"at Regensburg, Regensburg, Germany}
\author{Tim Leonhardt}%
\affiliation{JARA-FIT Institute for Quantum Information, Forschungszentrum J\"ulich GmbH and RWTH Aachen University, Aachen, Germany}
\author{Kentarou Sawano}%
\affiliation{Advanced Research Laboratories, Tokyo City University, Tokyo, Japan}
\author{Helge Riemann}%
\affiliation{Leibniz-Institut für Kristallzüchtung (IKZ), Berlin, Germany}
\author{Nikolay V. Abrosimov}%
\affiliation{Leibniz-Institut für Kristallzüchtung (IKZ), Berlin, Germany}
\author{Dominique Bougeard}%
\affiliation{Institut für Experimentelle und Angewandte Physik, Universit\"at Regensburg, Regensburg, Germany}
\author{Lars R. Schreiber}%
\email[email to: ]{lars.schreiber@physik.rwth-aachen.de}
\affiliation{JARA-FIT Institute for Quantum Information, Forschungszentrum J\"ulich GmbH and RWTH Aachen University, Aachen, Germany}

\date{\today}% It is always \today, today,
            
\begin{abstract}
Valley splitting is a key figure of silicon-based spin qubits. Quantum dots in Si/SiGe heterostructures reportedly suffer from a relatively low valley splitting, limiting the operation temperature and the scalability of such qubit devices. Here, we demonstrate a robust and large valley splitting exceeding 200\,$\upmu$eV in a gate-defined single quantum dot, hosted in molecular-beam epitaxy-grown $^{28}$Si/SiGe. The valley splitting is monotonically and reproducibly tunable up to 15\,\% by gate voltages, originating from a 6\,nm lateral displacement of the quantum dot. We observe static spin relaxation times $T_1>1$\,s at low magnetic fields in our device containing an integrated nanomagnet. At higher magnetic fields, $T_1$ is limited by the valley hotspot and by phonon noise coupling to intrinsic and artificial spin-orbit coupling, including phonon bottlenecking.
\end{abstract}

\maketitle

                              %displ
%\tableofcontents

\section{Introduction}
Silicon has proven to be an excellent host material for spin qubits \cite{Zwanenburg2013}. Demonstrated fidelities of single-qubit gates higher than 99.9\,\% \cite{Veldhorst2014,Laucht2015,Muhonen2015, Yoneda2017} are beyond the error correction threshold \cite{Fowler2012} and the fidelity of two-qubit gates is steadily increasing \cite{Veldhorst2015,Watson2017, Huang2019}. Nuclear purification \cite{Wild2012}, which reduces spin dephasing induced by hyperfine coupling, rendered this progress possible. Applying industrial fabrication processes and integrating conventional silicon electronics open up the perspective of a highly scalable and dense quantum computing architecture \cite{Hile2015,Ogorman2016,Veldhorst2017}. In particular, electrons trapped in electrostatically defined quantum dots (QDs) in Si/SiGe stand out by the excellent control of both the QD energies and the tunnel barriers \cite{Zajac2016, Mills2019}, both of which are important for universal multi-qubit manipulation. The SiGe barrier layer separates charged defects at the gate dielectric oxide interface \cite{Zajac2016,Sigillito2019} from the qubits in the silicon quantum well and thus reduces qubit dephasing due to charge noise. The major challenge for scaling up in this material system is its hardly controllable and reportedly small valley splitting $E_{\text{VS}}$, majoritarily below $70\,\upmu\text{eV}$ \cite{Borselli2011,Shi2011, Kawakami2014,Scarlino2017,Zajac2015,Mi2017,Watson2017,Ferdous2018,Mi2018-2,Borjans2018}. The excited valley state may then be occupied either by thermal excitation \cite{Kawakami2014,Vandersypen2016} or by fast spin relaxation \cite{Yang2013}, severely hampering the qubit control. $\Evs$ is known to crucially depend on the atomistic details at the Si/SiGe interface as well as on an applied out-of-plane electric field \cite{Ando1979,Boykin2004,Friesen2006, Friesen2007,Culcer2010,Shi2011,Yang2013}. A large $E_{\text{VS}}$ compared to the electron temperature and the Zeeman energy is thus highly desired to harness the full potential of Si/SiGe as one of the most promising hosts for spin qubits.
\par In this work, we present a singly charged gate-defined quantum dot with an integrated nanomagnet \cite{Petersen2013} in a molecular beam epitaxy (MBE)-grown $^{28}$Si/SiGe heterostructure, revealing a remarkably high valley splitting $\Evs$ beyond $200\,\upmu\text{eV}$. The energy values extracted both from pulsed gate spectroscopy and from the magneto-dependence of the spin relaxation time $T_1$ \cite{Huang2014} are consistent. We find this valley splitting to be robust against lateral displacements of the QD, experimentally realised by adjusting the dot-defining gate voltages while maintaining the dot size and orbital energy constant. The QD characteristics, including $E_{\text{VS}}$, are reproducible for continuous displacements of the QD and also for more abrupt switching between QD positions, indicating that this single-electron dot is not limited by potential disorder. We demonstrate a monotoic tunability of $E_{\text{VS}}$ of at least $15\,\%$ to result from a gate-controlled and reproducible displacement of the QD relativ to $^{28}$Si/SiGe interface steps \cite{Saraiva2011}. Analysing the magneto-dependence of the spin relaxation time $T_1$, which reaches values beyond $1$\,s, we find consistent results for a few QD positions regarding the relaxation mechanisms acting at the prominent spin-valley-dominated $T_1$ hotspot as well as in the lower and higher energy regions \cite{Huang2014,Huang2014-2, Borjans2018,Petit2018}.

\begin{figure*}[t]
\centering
\includegraphics[width=\textwidth]{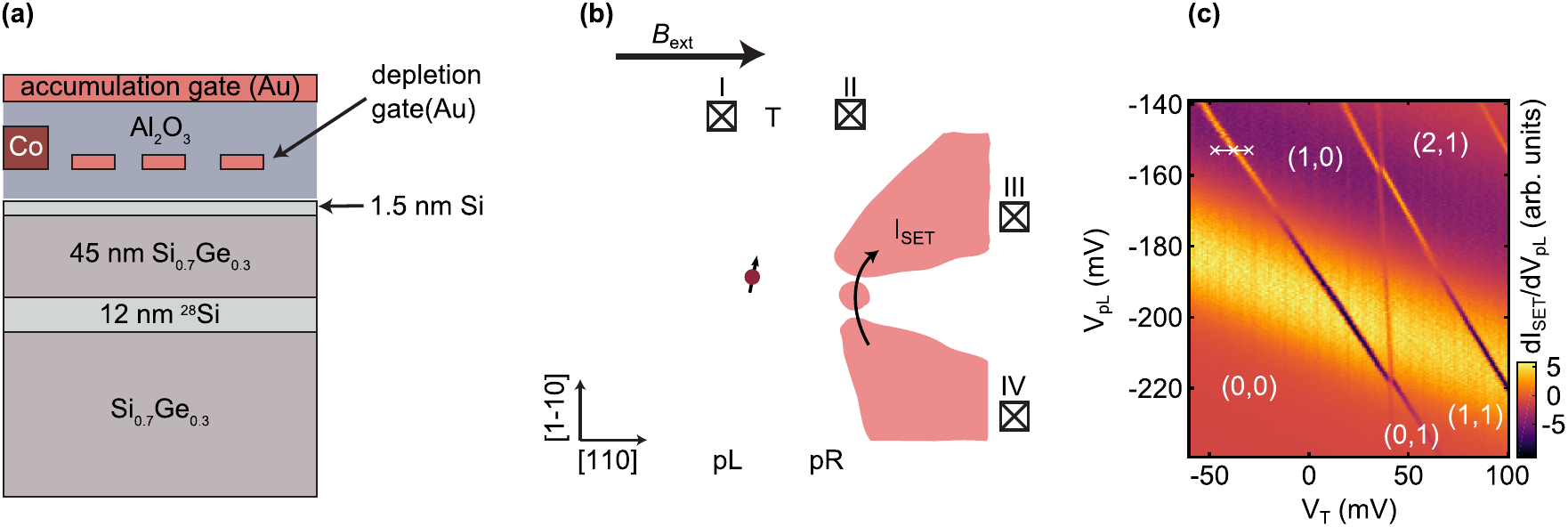}
\caption{\textbf{$^{28}$Si/SiGe device and charge stability diagram}. \textbf{(a)} Schematic cross section of the layer structure of the device. A 12\,nm purified $^{28}$Si/SiGe quantum well (with 60\,ppm $^{29}$Si) is grown on a Si$_{0.7}$Ge$_{0.3}$ virtual substrate. Electrostatic depletion gates (Au) are separated by 20\,nm Al$_2$O$_3$. A cobalt nanomagnet is added in this gate layer. \textbf{(b)} Top view: scanning electron image of the depletion gate layout of our DQD device with integrated nanomagnet and charge sensor. The global accumulation gate which is located in a different layer (see Supplementary Note 1 \cite{Supplementary}) is not resolved here. The electron reservoirs are labelled by squares and roman numbers. The cobalt nanomagnet indicated in blue introduces a magnetic field gradient and at the same time acts as an electrostatic confining gate. Crystallographic axes are indicated in the bottom left. The single-electron transistor formed on the right of the device for integrated charge sensing of the occupation of the QD is coloured in red. \textbf{(c)} Charge stability diagram of the DQD device close to the single-electron regime. The single-electron quantum dot (1,0) is solely tunnel-coupled to the reservoir I. Unload, load and read-out positions in pulsed operation are marked by white crosses.}
\label{Fig1}
\end{figure*}

\section{Device structure and setup}
Our double quantum dot (DQD) device with an integrated nanomagnet (Fig.~\ref{Fig1}) confines electrons in an undoped $^{28}$Si quantum well (QW) with 60~ppm residual $^{29}$Si. The heterostructure is fabricated by solid-source molecular beam epitaxy (MBE). A relaxed virtual substrate consists of a graded buffer grown at $500\degree\,\text{C}$ up to a composition of Si$_{0.7}$Ge$_{0.3}$ on a Si substrate without intentional miscut and a layer of constant composition Si$_{0.7}$Ge$_{0.3}$. It provides the basis for a 12\,nm $^{28}$Si quantum well (QW) grown using a source material of isotopically purified $^{28}$Si with 60~ppm of remaining $^{29}$Si with a rate of 0.14\,\AA/s at a substrate temperature of $350\degree\,\text{C}$. The QW is separated from the interface by 45\,nm Si$_{0.7}$Ge$_{0.3}$. The structure is protected by a 1.5\,nm naturally oxidised Si cap (see Fig.~\ref{Fig1}(a)). The implanted ohmic contacts to the QW are thermally activated by a rapid anneal at $700\degree\,\text{C}$. The mobility of the 2DEG formed in the QW, as obtained by Hall measurements at 1.5K temperature, is of the order of $55000\,\text{cm}^2/(\text{Vs})$ at an electron density of $6\times 10^{11}\,\text{cm}^{-2}$ and is limited by remote impurity scattering \cite{Wild2012}.\par
A layer of 20\,nm Al$_2$O$_3$ grown by atomic layer deposition insulates the depletion gate layer depicted in Fig.~\ref{Fig1}(b) and the underlying heterostructure. The depletion gates are fabricated by means of electron beam lithography. A Co nanomagnet, coloured in blue in Fig.~\ref{Fig1}(b), is added to the depletion gate layer in order to provide a local magnetic field gradient for electric dipole spin resonance (EDSR). A second gate layer, insulated from the depletion gates by 80\,nm of Al$_2$O$_3$, is used to induce a two-dimensional electron gas in the QW via the field effect and provide reservoirs for the dot-defining and charge-sensing parts of the device. A scanning electron micrograph of the second gate layer is shown in Supplementary Fig.~1 \cite{Supplementary}.
A single-electron transistor (SET) for charge detection is formed on the righ-hand side of the device between the reservoirs III and IV (coloured in red in Fig.~\ref{Fig1}(b)). The device was measured in an Oxford Triton dilution refrigerator at a base temperature of 40\,mK and electron temperature of 114\,mK. All DC lines are heavily filtered using pi-filters at room temperature followed by copper-powder filters and a second order $RC$ low-pass filter at base temperature (see Supplementary Note 1 \cite{Supplementary} for the details).

\section{Results}
\begin{figure}[tbp!]
\centering
\includegraphics{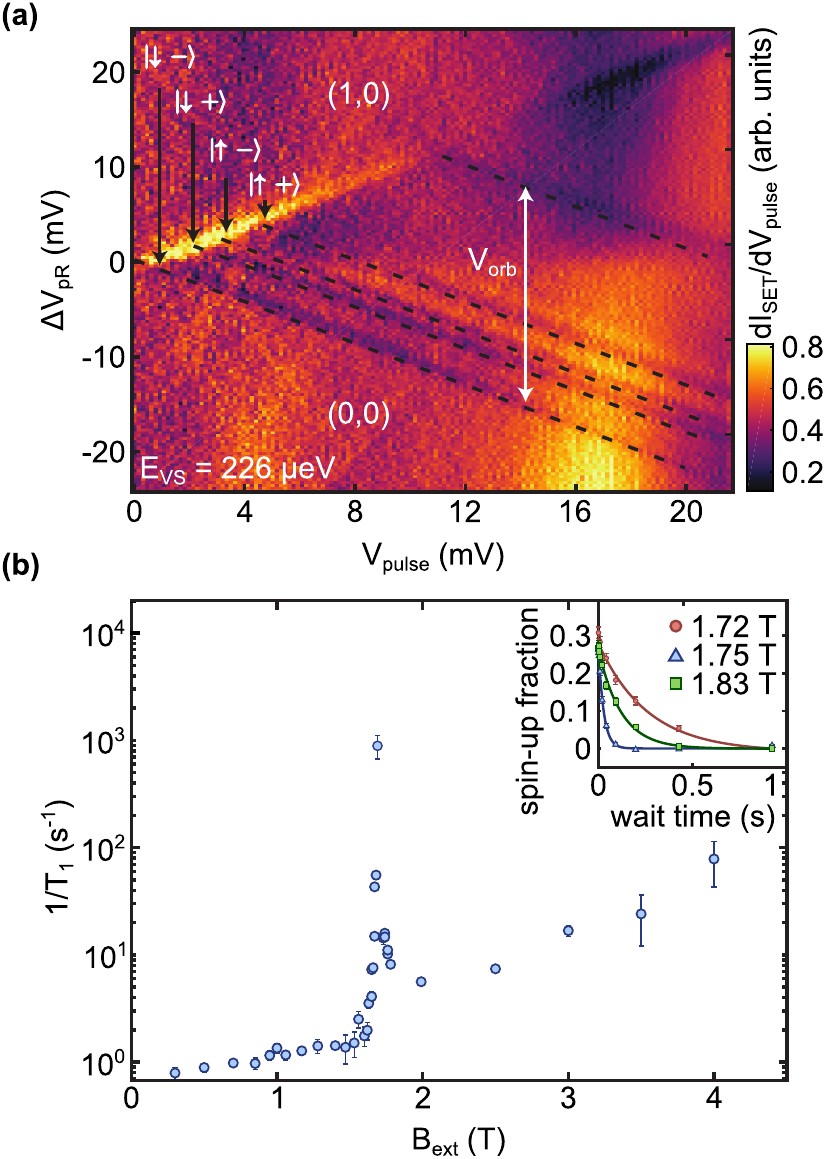}
\caption{\textbf{Pulsed gate spectroscopy and spin relaxation measurement.} \textbf{(a)} Pulsed gate spectroscopy on gate pR at 3\,T external magnetic field. We used the gate pR as it has a low cross-coupling to the reservoir tunnel-barrier. The time-averaged current change $\text{d}I_{\text{SET}}/\text{d}V_\text{pulse}$ is plotted as a function of the DC voltage offset $\Delta V_{\text{pR}}$ and the square pulse amplitude $V_{\text{pulse}}$, both applied to the gate pR. We chose $\Delta V_{\text{pR}}$ to be zero at the (0,0)-(0,1) dot-occupation transition. The plotted SET current is the sum of the SET currents from two measurements recorded at a pulse frequency of 10\,kHz and 50\,kHz, respectively. White labels mark the occupation of excited spin ($\ket{\downarrow}$, $\ket{\uparrow}$), valley ($\ket{-}, \ket{+}$) and orbital states. \textbf{(b)} Spin relaxation rate $1/T_1$ as a function of the external magnetic field $B_\text{ext}$. At $B\sim 2 \unit{T}$, we observe a steep rise in the spin relaxation (hotspot). Inset: Spin-up fraction as a function of the wait time for three magnetic field configurations in close proximity to the hotspot. The voltage tuning of the QD in \textbf{(b)} slightly differs from the configuration in \textbf{(a)}.}
\label{fig2}
\end{figure}

We have tuned the DQD to the single-electron regime as illustrated in the charge stability diagram in Fig.~\ref{Fig1}(c), where ($N_L$,$N_R$) denotes the electron occupancy in the left and right QD, respectively. In the following, the device is operated across the (0,0)-(1,0) charge transition far away from the (1,0)-(1,1) transition. An in-plane external magnetic field $B_{\text{ext}}$ is applied along the long axis of the nanomagnet as sketched in Fig.~\ref{Fig1}(b). The stray-field of the nanomagnet adds an additional longitudinal field component of $B_0= 40.71$\,mT at the left QD position, as shown in Supplementary Note 2 \cite{Supplementary}.

\subsection{Measurement of the orbital and valley splitting}

As a first characterisation of the device, we performed pulsed gate spectroscopy at the (0,0)-(1,0) charge transition. A 50\,\% duty-cycle square pulse with amplitude $V_{\text{pulse}}$ was added to the DC voltage of the gate pR indicated in Fig.~\ref{Fig1}(b) (see Supplementary Note 3 \cite{Supplementary}). As soon as $V_{\text{pulse}}$ is high enough to load an electron into an excited state, the effective loading rate increases: the average dot occupation and with it the average sensor current $I_{\text{SET}}$ changes. To clearly separate the spin states of this dot configuration, we fix the external magnetic field to 3\,T. The visibility of each transition depends on the ratio of the tunnel rate into the corresponding excited state and the frequency of the square pulse.
All four spin-valley states ($\ket{\downarrow-}, \ket{\downarrow +}, \ket{\uparrow-}, \ket{\uparrow+}$ and the excited orbital state are clearly resolved in Fig.~\ref{fig2}(a). The orbital spin states are well separated from the lower-lying spin-valley states. Using the applied Zeeman energy to determine the lever arm $\alpha = 0.06\,\text{eV}/\text{V}$ of gate pR, we calibrate the pR voltage to the change of chemical potential of the QD. This allows us to extract an orbital splitting of $E_\text{orb} = \alpha V_\text{orb} = 1.45\,\text{meV}$, defining a large window for the operation of spin qubits in the valley states. There, whenever the valley splitting $E_{\text{VS}}$ is of the order of the Zeeman splitting $E_\text{Z}$, spin-valley mixing becomes the dominant decay channel \cite{Huang2014}, the spin relaxation time $T_1$ then being ultimately limited by the inter-valley transition decay for $E_{\text{VS}} = E_\text{Z}$. In contrast to strongly confined electrons in MOS quantum dots \cite{Yang2013,Lai2011,Yang2019} this effect can hamper Pauli spin blockade read-out and operation at elevated temperature, thus affecting the scalability of Si/SiGe \cite{Vandersypen2016}. Here, from the pulsed gate spectroscopy, based on the lever arm, we estimate the valley splitting in our device to be extremely large with $E_{\text{VS}} \approx 226\,\upmu\text{eV}$ (two to three times larger than most of the reported values \cite{Kawakami2014,Scarlino2017,Zajac2015,Mi2017,Watson2017, Ferdous2018,Borjans2018}), motivating a more precise quantification of this valley splitting in the following.

To do so, we measure the spin relaxation rate in a large range of externally applied magnetic fields, allowing us to realise the condition $E_{\text{VS}} = E_\text{Z}$, where a peak (hotspot) in the spin relaxation is expected \cite{Huang2014,Borjans2018} due to spin-valley mixing. The position of this distinct and sharp peak of the spin relaxation spectrum allows a precise measure of $E_{\text{VS}}$. Compared to the pulsed gate spectroscopy experiments, we reduced the tunnel coupling to the reservoir to 5\,ms, in order to perform single-shot spin detection using energy-selective tunneling to the reservoir \cite{Elzerman2004}. This slightly reduces the QD size, leading to an increased orbital energy as will be discussed in the next section. We enhance the QD loading rate (see Supplementary Note 4 \cite{Supplementary}) by using the gate T shown in Fig.~\ref{Fig1}(c). We pulse the voltage of this gate to load and unload an electron between reservoir I and the left QD. We observe single exponential decays of the measured spin-up fraction as a function of the duration of the load pulse, as exemplarily shown in the inset of Fig.~\ref{fig2}(b) for three magnetic field values. The spin relaxation rate $1/T_{1}$ is extracted from these decays and plotted as a function of $B_{\text{ext}}$ in the main graph of Fig.~\ref{fig2}(b). The low magnetic field range, relevant for spin qubit operation ($B_{\text{ext}} \sim 0.7 $\,T for 20\,GHz electron spin resonance), yields long relaxation times $T_1>1$\,s which are rather insensitive to $B_{\text{ext}}$. We will address this regime together with the high field range in a later section and firstly concentrate on the sharp relaxation peak at $B_{\text{ext}} = 1.69$\,T which corresponds to the realisation of $E_{\text{VS}} = E_\text{Z}$. Taking the longitudinal magnetic field offset (40.71\,mT) of the nanomagnet at the QD position into account, the magnetic field position of this hotspot confirms the valley splitting in our device to be remarkably high, yielding a value of $E_{\text{VS}}=201\,\upmu\text{eV}$. Given the spin-selective nature of our readout, a variation of the amplitude of the loading pulse discussed in the Supplementary Note 5 \cite{Supplementary} validates our assignment of the spin states shown in Fig.~\ref{fig2}(a) and $E_{\text{VS}}=201\,\upmu\text{eV}$. We also exclude a valley splitting smaller than the thermal energy, since we observe a clear single-transition Chevron pattern (see Supplementary Note 6 \cite{Supplementary}). 

The energy splitting $E_{\text{VS}}$ between the two valley states is predicted to mainly depend on two parameters in Si/SiGe: the applied out-of-plane electric field $E_z$ and the sharpness of the QW interface \cite{Boykin2004, Friesen2007,Ando1979,Culcer2010}. In particular, an overlap of the electron wavefunction with atomic steps formed at this interface during the epitaxy of the heterostructure will reduce the valley splitting. Therefore, measuring a higher $E_{\text{VS}}$ for comparable electric field values $E_z$ may be a consequence of either a sharp QW interface or a small overlap, for example because of a strong lateral QD confinement. The latter may originate from potential disorder induced by bulk defects or defects at the oxide interface. Such defects will localise the QD uncontrollably. Note that the observed $E_\text{orb}=1.45\,\text{meV}$ (see Fig.~\ref{fig2}(a)) in our device points towards a QD potential which is formed by the potential of the gates. Assuming a lateral harmonic potential and taking the radius of the dot to be half of the full-width-at-half-maximum of the ground state, the corresponding QD has a radius of $\sim 19$\,nm. It thus overlaps with a large QW interface area (comparable to e.g. Ref. \cite{Zajac2016}, while in Ref. \cite{Borselli2011} a high $E_{\text{VS}}$ due to an unconventional small QD has been reported) and may thus indicate advantageous QW interface characteristics with regard to $E_{\text{VS}}$. The $^{28}$Si/SiGe heterostructure used here (see Methods) is grown by solid-source molecular beam epitaxy (MBE). The substrate temperature of $350\degree\,\text{C}$ in conjunction with adapted growth rates and the solid-source purity \cite{Wild2012} may be comparatively beneficial for achieving sharp interfaces. In order to evaluate the role of the out-of-plane electric field $E_z$ and to further exclude potential disorder as the reason for the observed high $E_{\text{VS}}$, we test in the following the evolution of $E_{\text{VS}}$ when varying the dot-defining gate voltages.
 \begin{figure}[tb!]
\centering
\includegraphics{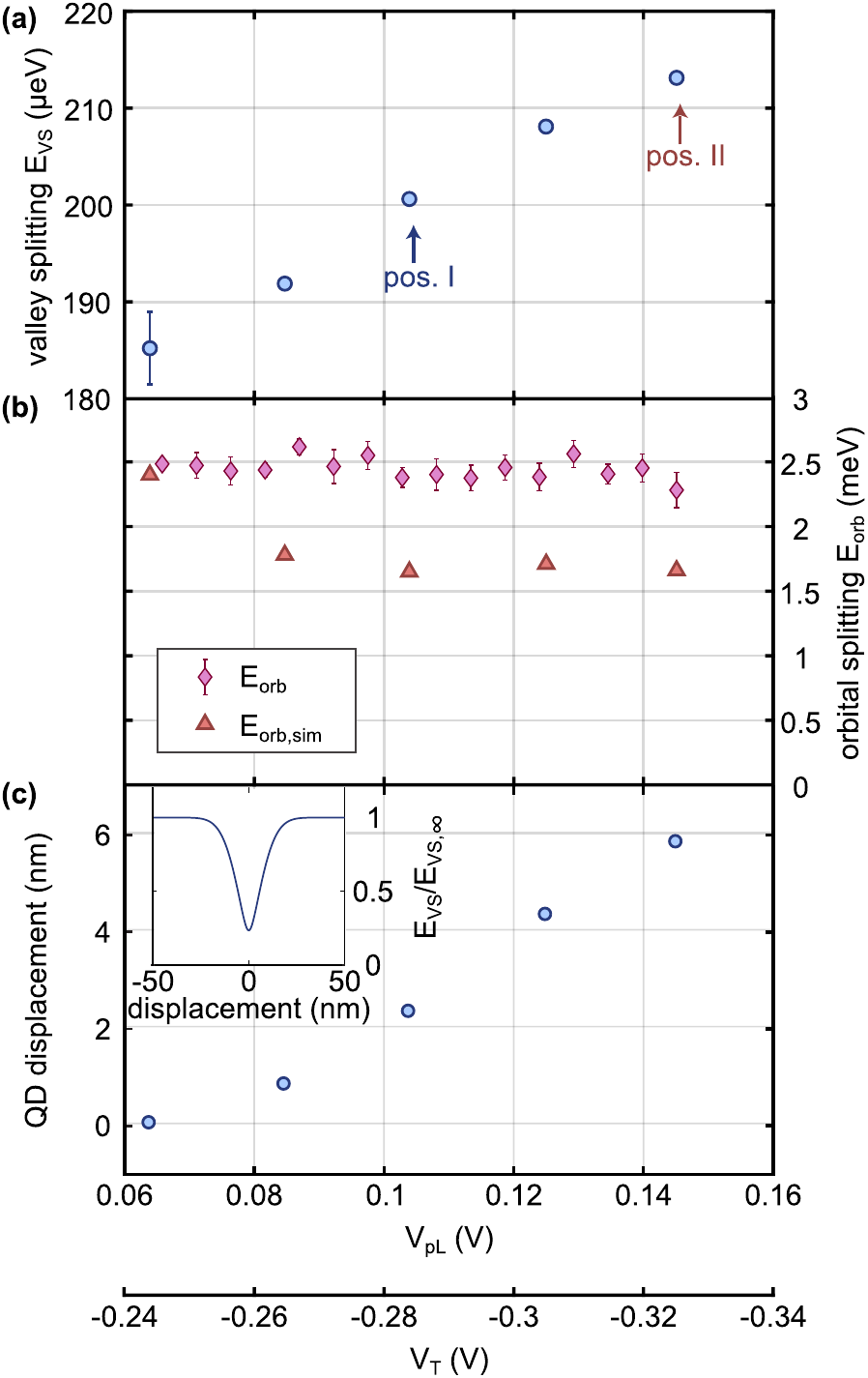}
\caption{\textbf{Control of the valley splitting.} \textbf{(a)} Valley splitting $E_{\text{VS}}$ for different QD configurations defined by the voltages on pL and T. The valley splitting $E_\text{VS}$ is extracted from the relaxation hotspot. The two labeled positions I and II correspond to the configurations used to analyse the spin relaxation mechanism in Fig.~\ref{fig5}. \textbf{(b)} Orbital splitting $E_\text{orb}$ extracted from pulsed gate spectroscopy measurements. $E_\text{orb,sim}$ indicates simulation results from a realistic Schrödinger-Poisson simulation of the single-electron QD (see Supplementary Note 7 \cite{Supplementary}).  \textbf{(c)} Simulated dot displacement for the given pL/T voltage configurations. Inset: Calculated change of valley splitting~$E_\text{VS}/E_{\text{VS},\infty}$ for a QD displacement across a single atomic step placed at 0\,nm.}
\label{fig4}
\end{figure}

\subsection{Tuning the valley splitting}
We systematically varied the voltages on both gates pL and T in steps of $20$\,mV, which keeps the chemical potential of the QD and the tunnel rate to the reservoir constant. For each voltage step, we measured the relaxation spectrum as a function of magnetic field in close proximity to the spin relaxation hotspot. The resulting $E_{\text{VS}}$ is depicted in Fig.~\ref{fig4}(a) as a function of the pL-T voltage configuration. The values of $E_{\text{VS}}$ remain remarkably high at the different gate voltages. All configurations were found to be robust and reproducible: the respective $E_{\text{VS}}$ values are also stable when jumping between different voltage configurations instead of steadily changing the voltages. Notably, we can monotonically tune $E_{\text{VS}}$ with pL/T: Changing the gate voltages in a range of $80\,\text{mV}$, $E_{\text{VS}}$ monotoically changes by $28\,\upmu\text{eV}$, corresponding to a tuning range of 15\,\%. Dependence of the singlet-triplet splitting on lateral QD displacement was observed by magnetospectroscopy in Ref. \cite{Shi2011}. 

To verify the impact of the gate voltage tuning on the lateral confinement potential of our single-electron QD, we measured the orbital splitting $E_\text{orb}$ by pulsed gate spectroscopy for all the voltage configurations shown in Fig.~\ref{fig4}(b): We find $E_\text{orb,exp}$ to remain unchanged at approximately 2.5\,meV. This trend of an orbital splitting unaffected by the gate voltage tuning is confirmed in a realistic Schrödinger-Poisson simulation of the single-electron QD which we subject to the same gate voltage tuning range. As shown in Fig.~\ref{fig4}(b), $E_\text{orb,sim}$ is also unchanged across this tuning range. Concomitantly, the in-plane shape of the QD remains constant in the simulation (see Supplementary Note 7 \cite{Supplementary}). We thus conclude that the in-plane confinement and size of our QD are conserved during the operated gate voltage variations. Note that 
we intentionally increased the tunnel-coupling to the reservoir for the pulsed gate spectroscopy shown in Fig.~\ref{fig2}(a) by increasing the size of the QD. Thus, we used a gate voltage-configuration, which is not covered by the configurations shown in Fig.~\ref{fig4}, but still we observe an $25\,\upmu\text{eV}$ larger valley splitting compared to the hotspot in Fig.~\ref{fig2}(b). This suggests a robust and tunable valley splitting in an even larger gate voltage parameter-space.

The impact of the out-of-plane electric field $E_z$ on the value of the valley splitting $E_{\text{VS}}$ in silicon qubit devices \cite{Friesen2007, Boykin2004,Saraiva2011,Saraiva2009} is particularly visible for MOS-based devices where the electron is strongly confined to the Si/SiO$_2$ interface. In such devices, a tunable valley splitting which depends linearly on $E_z$ for strong fields has been observed \cite{Yang2013}. Here, for our device, however, we find a comparatively weak value of the out-of-plane electric field $E_z$ in our simulation. $E_z$ also shows no significant variation between the different tested gate voltage configurations.
In fact, as shown in Fig.~\ref{fig4}(c), the main effect of our voltage tuning is a monotonic 6~nm in-plane displacement of the QD over the full conducted pL/T voltage range, according to the simulation. Noting the monotonic evolution of $E_{\text{VS}}$ when continuously adjusting the gate voltages on the one hand, combined to the observation of an excellent reproducibility of high $E_{\text{VS}}$ when more abruptly switching between the different gate voltage configurations on the other hand, we conclude that our device is not limited by potential disorder in the single-electron regime. Instead, our results strongly indicate that we can robustly and significantly displace the QD in the $^{28}$Si QW plane and that the valley splitting does not show abrupt disorder-induced variations on short lateral length scales.

Regarding the continuous variation of up to 15\,\% for $E_{\text{VS}}$ for the tested gate voltage configurations, we believe that it results from the controlled displacement relative to atomic steps at the $^{28}$Si/SiGe interface. We have calculated the change in valley splitting induced by an atomic step analytically, as shown in the inset of Fig.~\ref{fig4}(c). A single interface step leads to a valley phase of $\theta = 2k_0\cdot a_\text{Si}/4 = 0.85\pi$, where $a_\text{Si}/4$ is the height of a mono-atomic layer in silicon and $k_0 = 2\pi/a_\text{Si}\cdot0.85$ is the position of the valley minimum along the $\Delta$-direction. As can be seen in the inset of Fig.~\ref{fig4}(c), in the vicinity of an atomic step (placed at 0 nm), $\Evs$ changes by several tens of percent within a few nanometers, dropping down to $E_{\text{VS},\infty} \cos(\theta/2)$, where $E_{\text{VS},\infty}$ is the valley splitting far away from the interface step. 
%Changing the electron probability function across the step describes the observed shift in $E_{\text{VS}}$ \cite{Tariq2019}.  

\par 
\begin{figure*}[tpb!]
\centering
\includegraphics{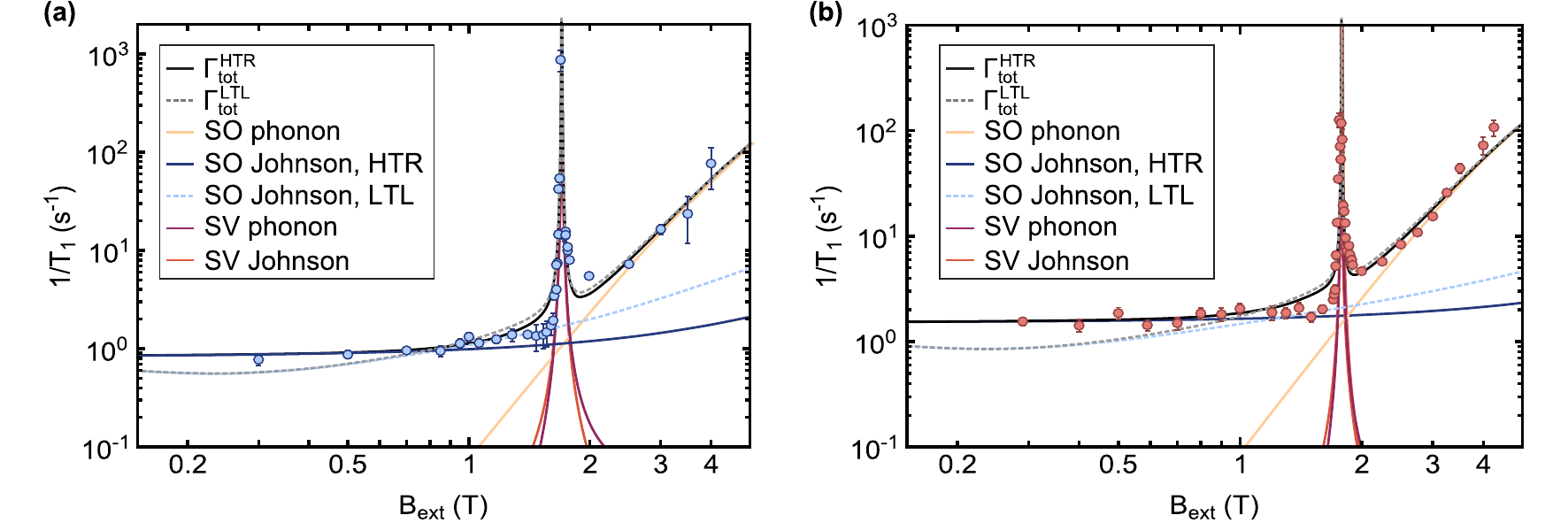}
\caption{\textbf{Spin relaxation rate fitted with rate equation. } \textbf{(a)} and \textbf{(b)} correspond to the position I and II, respectively as labeled in Fig.~\ref{fig4}. $\Gamma_\text{tot}^\text{HTR}$ corresponds to the elevated temperature resistor model and $\Gamma_\text{tot}^\text{LTL}$ to the lossy transmission line model, labeled by the solid black and grey dashed lines, respectively, see main text and Supplementary Note 8 \cite{Supplementary}. The different single relaxation contributions are indicated by the coloured lines. }
\label{fig5}
\end{figure*}
\subsection{Spin relaxation mechanisms}
Having analysed the high and robust valley splitting in our device, which manifests through the prominent spin relaxation hotspot in Fig.~\ref{fig2}(b), we now discuss the magnetic field-dependence of the spin relaxation time $T_1$ in more detail. The main three features of this dependence shown in Fig.~\ref{fig2}(b) are the sharp valley mixing hotspot, a strong $B_{\text{ext}}$-dependence in the field regime beyond the hotspot and a roughly magnetic field-independent $T_1$ in the low-field regime. We have fitted these magnetic field-dependencies for two gate voltage configurations of the device in a large range of magnetic fields (configurations I and II, marked in Fig.~\ref{fig4}). The fitting formula is based on a rate equation including spin-valley (SV) and both intrinsic and artificial spin-orbit (SO) coupling, the latter of which is given by the simulated gradient magnetic field of the nanomagnet. These coupling mechanisms are combined with Johnson noise (J) and phonon noise (ph), respectively \cite{Huang2014, Petit2018, Borjans2018}:  
\begin{equation}
    \Gamma_\text{tot} = \Gamma_\text{SV,J}+\Gamma_\text{SO,J}+\Gamma_\text{SV,ph} +\Gamma_\text{SO,ph}
    \label{eq:gamma_total}
\end{equation}
Our approach to the rate equation and the extracted fit parameters for two different QD positions are discussed in more detail in the Supplementary Note 8 \cite{Supplementary}.
The coloured solid lines in Fig.~\ref{fig5} represent the different spin relaxation contributions $\Gamma_\text{i}$ of Eq. \ref{eq:gamma_total} in both gate configurations, allowing us to analyse the contributions of the relevant relaxation mechanisms to the different magnetic field regions. The prominent spin-relaxation hotspot is in excellent agreement with a fit dominated by SV combined with both phonon and Johnson noise ($\Gamma_\text{SV,J}+\Gamma_\text{SV,ph}$). From the peak width we determine a value of the product $\Delta r^{+-}$ of $\approx 15.1$\,neV$\cdot$nm for the configuration in Fig.~\ref{fig5}(a), where $\Delta$ and $r^{+-}$ are the spin-valley coupling energy and the valley-orbit dipole strength, respectively. In the Supplementary Note 8 \cite{Supplementary}, we also evaluated the peak width which would result if the hotspot was of orbital nature: we find a significantly larger value than the one extracted from Fig.~\ref{fig5} in both dot configurations, strongly supporting our attribution of the spin relaxation hotspots to the spin-valley coupling.

The rate equation suggests the magnetic field range beyond the hotspot to be governed by SO and phonon noise. The intrinsic and artificial SO interactions contribute with magnetic field dependencies of $(B_{\text{ext}})$$^7$ and $(B_{\text{ext}})$$^5$, respectively. In Fig.~\ref{fig5}, we find the best fit agreement when combining both dependencies with a phonon bottleneck for coupling to a QD \cite{Meunier2007, Huang2014-2}. The bottleneck effect gets relevant beyond $2.4\,\text{T}$, corresponding to $\sim 67\,\text{GHz}$, the point where the inverse of the phonon wave number exceeds the QD size. 

In the low-field regime, we observe a remarkably long spin relaxation rate ($T_1>1$\,s) (Fig.~\ref{fig5}(a)), which is nearly independent from $B_{\text{ext}}$. This saturation may appear also in Ref. \cite{Borjans2018}, however, overlayed by the low energy side of the valley relaxation hotspot, due to a lower $E_{\text{VS}}$ in the devices discussed there. We have considered two mechanisms as possible sources for the experimentally observed trend. Firstly, capacitively coupling artificial SO combined with white electric noise (e.g. Johnson noise from a high temperature resistor (HTR) at a temperature above 2.5\,K) to the QD will result in a $B_{\text{ext}}$-independent spin relaxation. However, as shown in the Supplementary Note 8 \cite{Supplementary}, we find the magnitude of the fitted Johnson noise (blue solid line in Fig.~\ref{fig5}) to be much larger than the actual resistor values at elevated temperature in our setup, the high-frequency Johnson noise of which is attenuated by copper-powder filters and high-frequency attenuators, respectively. Given this discrepancy with the characteristics of our setup we find this mechanism solely inducing our observed $T_1$-trend unlikely, although the fit shown in Fig.~\ref{fig5} may look reasonable.

A second candidate for inducing Johnson noise into the device are fluctuations in the electron reservoir coupled to the QD. As discussed in more detail in the Supplementary Note 8 \cite{Supplementary}, we therefore considered the ohmic contact resistances and resistances of the Si/SiGe 2DEG, to form a lossy transmission line (LTL) together with the global inducing gate (blue dashed-line in Fig.~\ref{fig5}). A spin relaxation which includes this type of Johnson-noise on the electron reservoirs combined with $\Gamma_\text{SV,J}+\Gamma_\text{SV,ph} +\Gamma_\text{SO,ph}$ is plotted as the grey dashed line in Fig.~\ref{fig5} ($\Gamma_\text{tot}$,LTL). Although the fit may look reasonable for the gate voltage configuration in Fig.~\ref{fig5}(a), we find it to be in less good agreement with the other tested QD position  where the low-field regime is clearly independent of the magnetic field (Fig.~\ref{fig5}(b)).
The black line shown in Fig.~\ref{fig5} ($\Gamma_\text{tot}$,HTR) depicts the total fit $\Gamma_\text{tot}$($B_{\text{ext}}$) to the measured spin relaxation rate, combining the dominant relaxation mechanisms for each magnetic field region for this exemplary gate configuration. All in all, we find an excellent agreement for all tested dot positions for the sharp hotspot peak and the high-field region, while more statistics and a higher magnetic field resolution of the low-field region seem to be required to unambiguously identify the dominant noise source.

\section{Discussion}
In conclusion, we have characterised a single-electron spin qubit in MBE-grown isotopically purified Si/SiGe with pulsed gate spectroscopy and through its magneto-dependence of the spin relaxation time $T_1$. The dot-defining gate layout allows us to laterally displace this QD (radius $\sim 19$\,nm) by changing the gate voltages while keeping its size and orbital energy $E_\text{orb,exp}$ constant. We find consistently high valley splittings $E_{\text{VS}}\sim 200\,\upmu\text{eV}$ and a well-separated $E_\text{orb,exp}$, both when continuously sweeping gate voltages between different QD positions and also when more abruptly switching between QD positions. The robustness and reproducibility of the QD characteristics with the dot displacement as well as our simulation of the out-of-plane electric field $E_z$ strongly indicate that our QD is not dominated by potential disorder and that the consistently high value of $E_{\text{VS}}$ does not result from disorder confinement or $E_z$, but seems to be due to  $^{28}$Si/SiGe QW interfaces that are advantageous in terms of $E_{\text{VS}}$, albeit the heterostructure undergoes an implantation anneal at $700\degree \,\text{C}$. One reason for these findings may be the comparatively low substrate temperatures used in MBE. The monotonic variation of $E_{\text{VS}}$ of 15\,\% which we observe experimentally is attributed to a reproducible and gate-controlled displacement relative to atomic steps at the QW interface. This robust displacement process supports our conclusion of dealing with a comparatively smooth (i.e. free from short-length-scale disorder) $^{28}$Si/SiGe interface for this device. The closer analysis of the magneto-dependence of the relaxation time $T_1$ confirms the prominent relaxation hotspot to be dominated by spin-valley coupling, combined with Johnson and phonon noise. At lower magnetic fields, the relaxation times $T_1>1$\,s are relatively insensitive to the external magnetic field. At higher fields, we find very good agreement for fits that include phonon noise acting with a combination of intrinsic and artificial spin-orbit coupling, in the presence of phonon bottlenecking. The observed artificial spin-orbit coupling is fully consistent with the field gradient induced by the nanomagnet which is integrated into the dot-defining gate layout. The demonstration of consistently high $E_{\text{VS}}$ and the absence of short-length-scale disorder in this device provides a promising perspective towards higher-yield spin qubit devices in Si/SiGe QDs.

\begin{acknowledgments}
The authors thank Dieter Weiss and Rupert Huber for technical support. This work has been funded by the German Research Foundation (DFG) within the projects BO 3140/4-1, 289786932 and the cluster of excellence "Matter and light for quantum computing" (ML4Q) as well as by the Federal Ministry of Education and Research under Contract No. FKZ: 13N14778. Project Si-QuBus received funding from the QuantERA ERA-NET Cofund in Quantum Technologies implemented within the European Union's Horizon 2020 Programme. 
\end{acknowledgments}

\nocite{Elzerman2004,Simmons2011,Yang2013,Huang2014-2,Huang2014,Golovach2004,vanVleck1940,Khaetskii2001,Petit2018,Yu2010,Tahan2014,Boross2016,Weiss2008,Pozar2005}

\end{document}